\documentclass{PoS}
\usepackage{amsmath} 
\usepackage{slashed}

\newcommand{\hmchipt}{HM$\chi$PT}

\title{The $B^*B\pi$ coupling with relativistic heavy quarks}

\ShortTitle{The $B^*B\pi$ coupling with relativistic heavy quarks}

\author{\speaker{B.~Samways}, J.M.~Flynn, C.T.~Sachrajda\\
  School of Physics and Astronomy, University of Southampton, Southampton, SO17 1BJ, UK}
\author{P.~Fritzsch\\
  Institut f\"{u}r Physik, Humboldt Universit\"{a}t, Newtonstr. 15, 12489 Berlin, Germany}
\author{T.~Kawanai, C.~Lehner\\
  Physics Department, Brookhaven National Laboratory, Upton, NY 11973, USA}
\author{R.S.~Van de Water\\
  Theoretical Physics Department, Fermi National Accelerator Laboratory, Batavia, IL 60510, USA}
\author{O.~Witzel\\
  Center for Computational Science, Boston University, 3 Cummington Mall, Boston, MA 02215, USA}

\abstract{We report on a calculation of the $B^*B\pi$ coupling in lattice QCD. The strong matrix element $\langle B \pi | B^*\rangle$ is directly related to the leading order low-energy constant in heavy meson chiral perturbation theory (\hmchipt{}) for $B$-mesons. We carry out our calculation directly at the $b$-quark mass using a non-perturbatively tuned clover action  that controls discretisation effects of order $|\vec{p}a|$ and $(ma)^n$ for all $n$. Our analysis is performed on RBC/UKQCD gauge configurations using domain wall fermions and the Iwasaki gauge action at two lattice spacings of $a^{-1}=1.73(3)$~GeV, $a^{-1}=2.28(3)$~GeV, and unitary pion masses down to 290~MeV. We achieve good statistical precision and control all systematic uncertainties, giving a final result for the \hmchipt{}  coupling $g_b = 0.569(48)_{stat}(59)_{sys}$ in the continuum and at the physical light-quark masses. This is the first calculation performed directly at the physical $b$-quark mass and lies in the region one would expect from carrying out an interpolation between previous results at the charm mass and at the static point.}

\FullConference{31st International Symposium on Lattice Field Theory LATTICE 2013\\
   July 29 - August 3, 2013\\
    Mainz, Germany}

\begin{document}

\section{Introduction}
The power of lattice QCD in probing of the Standard Model, and uncovering evidence for new physics, lies predominantly in the flavour sector. To constrain the CKM unitarity triangle there are many inputs required that can only be accessed non-perturbatively, particularly in the $B$-meson sector. For instance, lattice calculations of the decay constants $f_{B}$ and $f_{B_s}$ are necessary inputs for neutral $B$-meson mixing calculations and for the Standard Model predictions of BR$(B\to\tau\nu)$ and BR$(B_s\to\mu^+\mu^-)$ respectively. Furthermore, lattice calculations of the $B\to\pi l\nu$ form factor allow a determination of the CKM matrix element $|V_{ub}|$.  
For both semileptonic form factors and mixing matrix elements, the lattice precision lags behind experiment. However the experimental measurements will continue to improve with the large data sets available at Belle~II and the LHCb upgrade.  Therefore it is essential to further reduce the theoretical uncertainties in the non-perturbative hadronic parameters in order to maximise the scientific impact of current and future $B$-physics experiments.\\
A major source of uncertainties in all previous lattice calculations is from practical difficulties simulating at physical light-quark masses. Theoretical insight from \hmchipt{} can guide extrapolations down to the physical point, but lack of knowledge of the low-energy constants (LECs) of the theory introduces unwanted uncertainties. For example, at next-to-leading order (NLO) in  \hmchipt{} the dependence of $f_{B_d}$ on the light-quark (or equivalently, pion) mass is given by
\begin{equation}
  f_{B_d} = F\left(1+\frac{3}{4}(1+3g_b^2)\frac{M_{\pi}^2}{(4\pi f_{\pi})^2}\log(M_{\pi}^2/\mu^2)\right) + \cdots ,
\end{equation}
where $g_b$ is the leading order LEC of the theory, and is directly related to the strong coupling $g_{B^*B\pi}$. In this work we perform the first calculation of the coupling $g_b$ directly at the $b$-quark mass.

\section{Heavy Meson Chiral Perturbation Theory}\label{heavymesontheory}
In the infinite quark mass limit, symmetries predict that properties of heavy-light mesons will be independent of the heavy quark's spin and flavour quantum numbers. Combining this with the chiral symmetry present in the $m_q \rightarrow 0$ limit of QCD provides the basis for heavy meson chiral perturbation theory (\hmchipt{}). This effective theory of QCD is a joint expansion in powers of the inverse heavy-quark mass $1/m_h$ and the light-quark-mass $m_q$. At lowest order the interactions between the heavy and light mesons are determined by a Lagrangian with a single LEC~\cite{Wise1992}
\begin{equation}
  \label{eq:HMCHIPTLagrangian}
  \mathcal{L}_{HM\chi PT}^{int} = g Tr \left(\bar{H}_aH_b\mathcal{A}^{ba}_{\mu}\gamma^{\mu}\gamma{5}\right),
\end{equation}
with
\begin{equation}
  \mathcal{A}_{\mu} = \frac{i}{2}\left(\xi^{\dagger}\partial_{\mu}\xi + \xi \partial_{\mu}\xi^{\dagger} \right)
\end{equation}
and $\xi = \exp\left(i\mathcal{M}/f_{\pi} \right)$, where $\mathcal{M}$ represents the usual octect of pseudo-goldstone bosons. The coupling $g_b$ can be related to the coupling responsible for the strong decay $B^* \rightarrow B\pi$, defined as
\begin{equation}
  \label{eq:gBstarBpi}
  \langle B(p) \pi(q)|B^*(p^{\prime}, \lambda)\rangle = -g_{B^*B\pi}\:q\cdot\epsilon^{\lambda}(p^{\prime}).
\end{equation}
Equivalently, the same matrix element can be evaluated at leading order in \hmchipt{},
\begin{equation}
  \label{eq:gbstarbpihmchipt}
  \langle B(p) \pi(q)|B^*(p^{\prime}, \lambda)\rangle = -\frac{2M_B}{f_{\pi}} g_b q\cdot\epsilon^{\lambda}(p^{\prime}),
\end{equation}
giving the relationship
\begin{equation}
  \label{gbtogbstarbpi}
  g_{B^*B\pi} = \frac{2M_B}{f_{\pi}} g_b.
\end{equation}
Performing an LSZ reduction and using the partially-conserved axial current relation for a soft pion, Eq. (\ref{eq:gBstarBpi}) becomes
\begin{equation}\label{pionpole}
g_{B^*B \pi}\; q \cdot \epsilon^{\lambda}(p^{\prime}) = i q_{\mu} \frac{M_{\pi}^2-q^2}{f_{\pi}M_{\pi}^2}  \int d^4x\; e^{iq\cdot x} \langle B(p)|A_{\mu}(x)|B^*(p^{\prime}, \lambda)\rangle,
\end{equation}
where $A^\mu = \bar q \gamma^\mu \gamma_5 q$ is the light-quark axial vector current. If we parameterise the axial-current matrix element in terms of form factors

\begin{equation}\label{formfactors}
  \begin{split} 
    \langle B(p)| A^\mu | B^*(p^{\prime},\lambda) \rangle &= 2M_{B^*} A_0(q^2) \frac{\epsilon\cdot q}{q^2}q^\mu + (M_{B^*}+M_{B})A_1(q^2)\left[\epsilon^\mu - \frac{\epsilon\cdot q}{q^2}q^\mu\right] \\ 
    &+ A_2(q^2)\frac{\epsilon\cdot q}{M_{B^*}+M_{B}}
    \left[p^{\mu}+p^{\prime\mu} - \frac{M_{B^*}^2-M_{B}^2}{q^2} q^\mu
    \right],
  \end{split}
\end{equation}
we see that at $q^2=0$
\begin{equation}
  g_{B^*B\pi}=\frac{2M_{B^*}A_0(0)}{f_{\pi}}.
\end{equation}
On the lattice, we cannot simulate exactly at $q^2=0$ without using twisted boundary conditions. Furthermore, and from Eq.
(\ref{pionpole}) we see the form factor $A_0$ contains a pole at the pion mass,
so it will be difficult to extrapolate to $q^2=0$ in a controlled manner. However, the form factor
decomposition in Eq.~\eqref{formfactors} must be free of
nonphysical poles, which allows us to obtain the relation

\begin{equation}
  g_{B^*B\pi} = \frac1{f_\pi} \left[ (M_{B^*}+M_B)A_1(0) + (M_{B^*}-M_B)A_2(0)\right] .
\end{equation}

\section{Calculational strategy}
Our analysis is carried out using ensembles produced by the RBC and UKQCD collaborations~\cite{Aoki2010a} with the Iwasaki gauge action and 2+1 flavour dynamical domain-wall fermions. The configurations are at two lattice spacings, the finer $32^2$ ensembles have an inverse lattice spacing of $a^{-1} = 2.28(3)$~GeV and the coarser $24^3$ ensmbles have $a^{-1} = 1.73(3)$~GeV, corresponding to approximately 0.08~fm and 0.11~fm respectively. All ensembles have a spatial extent of 2.6~fm. We simulate with unitary light-quarks corresponding to pion masses down to $M_{\pi}=289$~MeV. On all ensembles, the sea strange-quark mass is tuned to within 10\% of its physical value. The fifth dimensional extent of both lattices is $L_S = 16$, corresponding to a residual quark mass of $(m_{\rm res}a)=0.003$ on the $24^3$ lattice and $(m_{\rm res}a)=0.0007$ on the $32^3$ lattice. Full details of the ensembles and propagators used are presented in Table~\ref{fig:ensembles}.\\
\begin{table}[!ht]
  \centering
    \begin{tabular}[c]{ccccccc}
      \hline
      $L/a$  & $a$(fm) & $m_la$ & $m_sa$ & \#Configs & \#Sources & $M_{\pi}$(MeV)\\
      \hline
      24 & 0.11  & 0.005 & 0.04  & 1636 & 1 & 329\\ 
      24 & 0.11  & 0.010 & 0.04  & 1419 & 1 & 422\\ 
      24 & 0.11  & 0.020 & 0.04  & 345 & 1 & 558\\ 
      32 & 0.08  & 0.004 & 0.03  & 628 & 2 & 289\\
      32 & 0.08  & 0.006 & 0.03  & 889 & 2 & 345\\
      32 & 0.08  & 0.008 & 0.03  & 544 & 2 & 394\\
      \hline
    \end{tabular}
  \caption{Lattice simulation  properties. All ensembles are generated using 2+1 flavours of domain-wall fermions and the Iwasaki gauge action. All valence pion masses are equal to the sea-pion mass.}
  \label{fig:ensembles}
\end{table}
In this work we use the Relativistic Heavy Quark (RHQ) action~\cite{El-Khadra1996a, Aoki2001, Christ2006} to simulate fully relativistic bottom quarks whilst controlling discretisation effects. The RHQ action is an anisotropic Wilson action with a Sheikholeslami-Wohlert term:
\begin{equation*}
  \label{eq:RHQ}
    S_{RHQ} = a^4\sum_{x,y} \bar{\psi}(y)\left(m_0+\gamma_0D_0+\xi\vec{\gamma}\cdot\vec{D}-\frac{a}{2}(D_0)^2-\frac{a}{2}\xi(\vec{D})^2+\sum_{\mu \nu}\frac{ia}{4}c_p\sigma_{\mu\nu}F_{\mu\nu}\right)_{y,x}\psi(x).
\end{equation*}
El Khadra, Kronfeld, and Mackenzie showed that for correctly tuned parameters the anisotropic Clover action can be used to describe heavy quarks with controlled cut-off effects to all orders in $ma$ and of $\mathcal{O}(|\vec{p}a|)$~\cite{El-Khadra1996a}. Christ, Li, and Lin~\cite{Christ2006} later showed that only three independent parameters need to be determined and, further, presented a method for performing this parameter tuning non-perturbatively~\cite{Lin2006}. This tuning has now been completed for $b$-quarks~\cite{tuning12} on the RBC/UKQCD configurations and these results are exploited in this calculation.  

\subsection{Ratios}
To access the matrix element in Eq. (\ref{formfactors}) we calculate the lattice three-point function:
\begin{equation}\label{threepoint}
C_{\mu \nu}^{(3)}\left( t_x, t_y; \bar{p}, \bar{p}^{\prime} \right) = \sum_{\bar{x}\bar{y}} e^{-\imath \bar{p}\cdot\bar{x}} e^{-\imath \bar{p}^{\prime}\cdot\bar{y}} \langle B(y) A_{\nu}(0) B^*(x) \rangle_{t_x<0<t_y}
\end{equation}
and the vector and pseudoscalar meson two point functions. If we set both the vector and pseudoscalar momenta to zero in Eq.~(\ref{threepoint}) we can see from Eq.~\eqref{formfactors} that the only form factor accessible is
$A_1$. Therefore we form the ratio:
\begin{equation}
  R_1 = \frac{C_{i,i}^{(3)}\left( t_x, t_y; \bar{p}=0, \bar{p}^{\prime}=0 \right)Z_{B}^{1/2}Z_{B^*}^{1/2}}{C_{BB}^{(2)}\left(t_y;\bar{p}=0\right) C_{B^*_{i}B^*_{i}}^{(2)}\left(T-t_x;\bar{p}=0\right) } = (M_{B^*}+M_B)A_1(q_0^2),
\end{equation}
where $Z_{B}$ and $Z_{B^*}$ are the amplitudes extracted from the pseudoscalar and vector two-point functions.\\
To access the other form factors we need to inject a unit of momentum,
such that $\bar{q} = \bar{p} = (1,0,0)\times2\pi/L$ and
$\bar{p}^{\prime}=0$. Following~\cite{Abada2002}, we
define further ratios $R_2$, $R_3$ and $R_4$ which allows access to the form factor $A_2$ through
\begin{equation} \label{A2/A1} \frac{A_2}{A_1} =
  \frac{(M_{B^*}+M_B)^2}{2M_B^2q_1^2}\left[-q^2_1+E_{B^*}(E_{B^*}-M_B)-\frac{M_{B^*}^2(E_{B^*}-M_B)}{E_{B^*}}\frac{R_3}{R_4}
    - i\frac{M_{B^*}^2q_1}{E_{B^*}}\frac{R_2}{R_4}\right].
\end{equation}
The ratio in Eq.~(\ref{A2/A1}) is obtained at non-zero values of
$q^2$ and needs to be extrapolated to $q^2=0$. However, its contribution is 
suppressed by the ratio $(M_{B^*} - M_{B})/(M_{B^*} + M_{B})$. The form factor
$A_1$ is obtained at $q^2 = (M_{B^*} - M_B)^2$, but examination shows that the slight extrapolation to $q^2=0$ is not necessary at the resolution possible with the available statistics. If we define functions $G_1$ and $G_2$ 
\begin{equation}
  \label{eq:G1G2}
  \begin{split}
  G_1(q^2) &= (M_{B^*}+M_B)A_1(q^2),\\
  G_2(q^2) &= (M_{B^*}-M_B) A_2(q^2),
  \end{split}
\end{equation}
we can write the coupling as $G_1(0)$ plus a small correction from the ratio $G_2/G_1$, giving
\begin{equation}
  \label{eg:gb}
  g_{b} = \frac{Z_A}{2M_B} G_1(0)\left( 1+\frac{G_2(0)}{G_1(0)} \right),
\end{equation}
where $Z_A$ is the light axial vector current renormalisation factor. We use the determination of $Z_A$ from the RBC/UKQCD combined analysis of the light hadron spectrum, pseudoscalar meson decay constants and quark masses on the $24^3$ and $32^3$ ensembles~\cite{Aoki2010a}.

\section{Results}
\begin{figure*}[!ht]
  \centering
  \begin{tabular}[c]{cc}
    \includegraphics[angle=0,width=0.47\linewidth ]{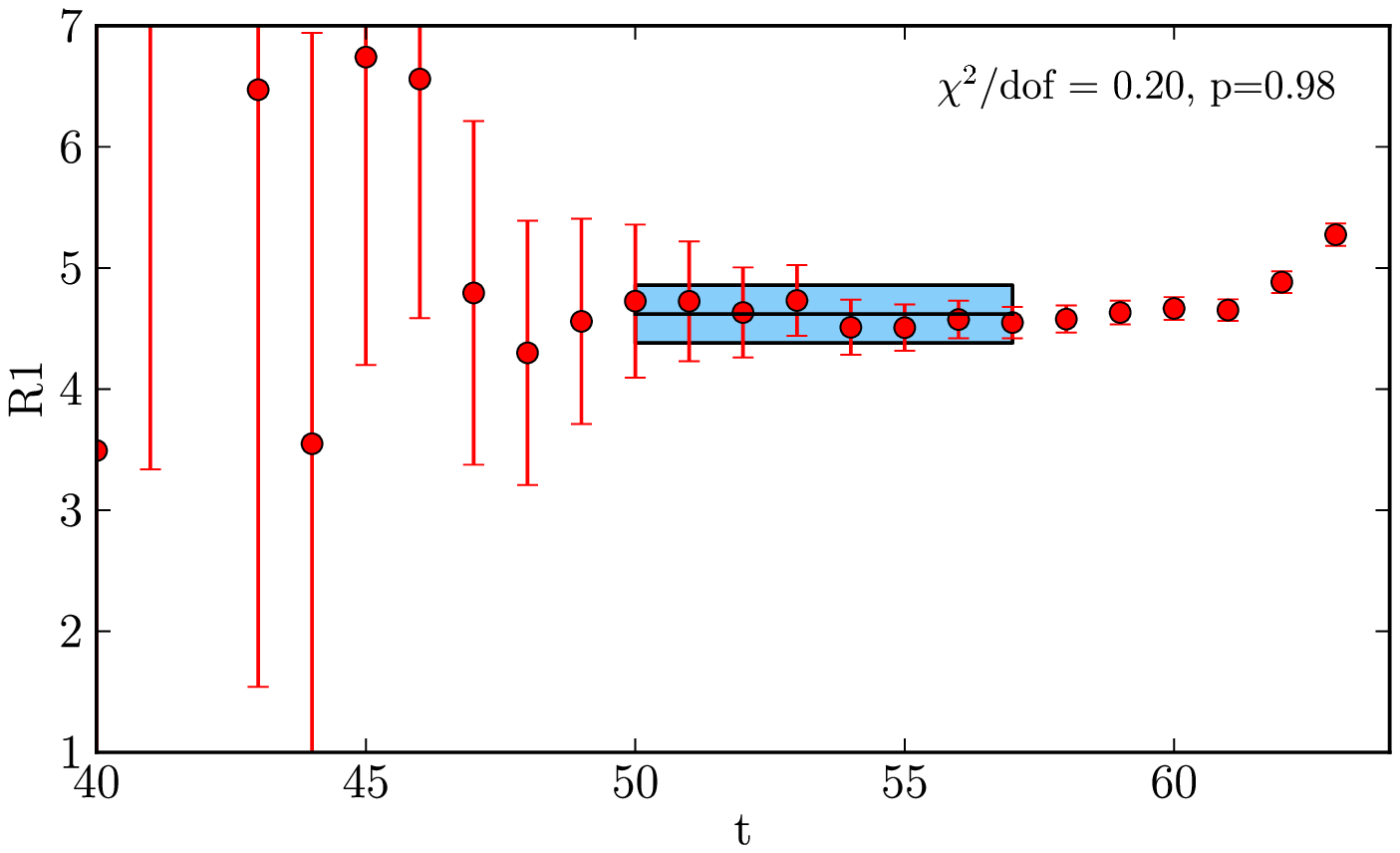}& \includegraphics[angle=0,width=0.47\linewidth ]{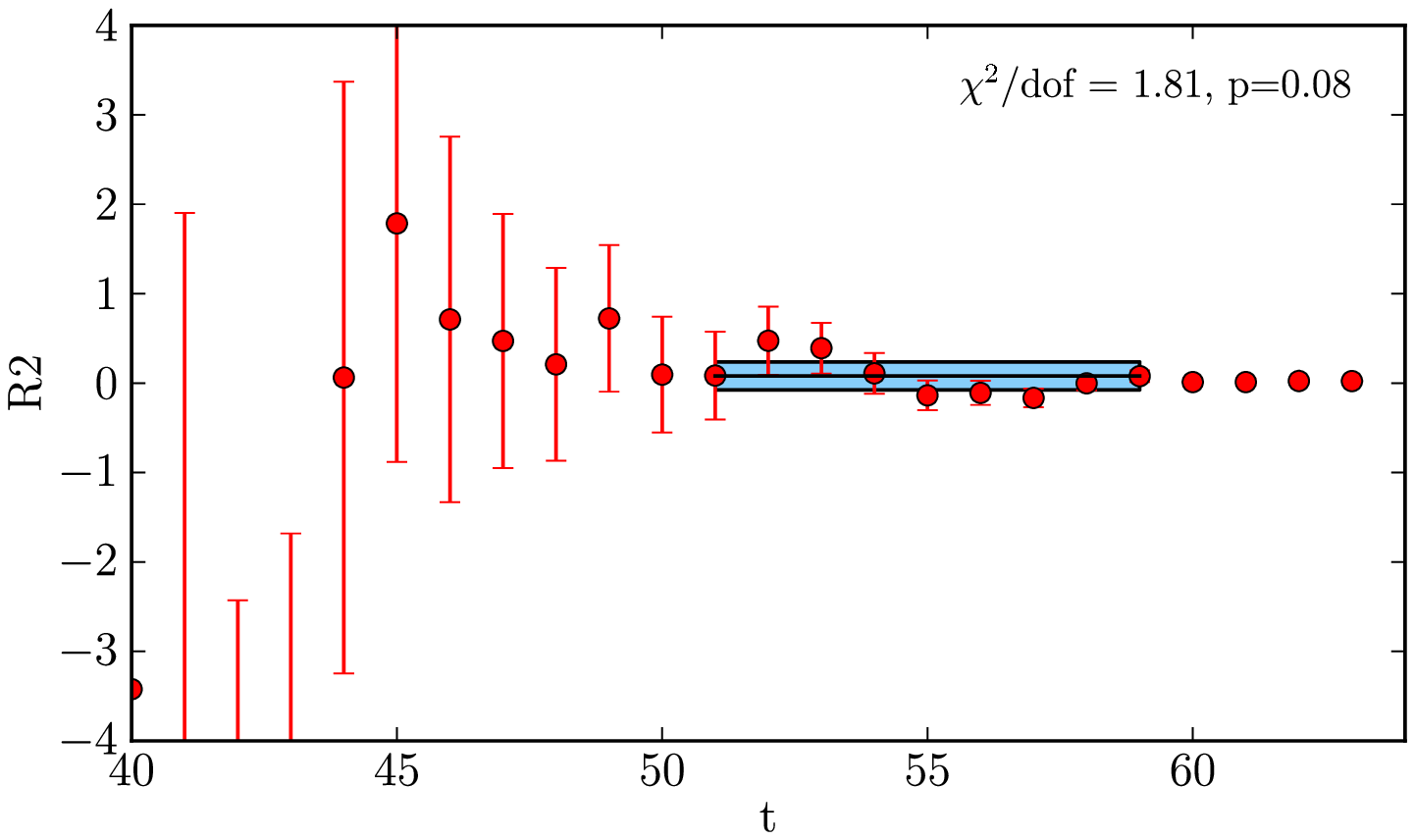}
  \end{tabular}
  \caption{Ratios $R_1$ (left), $R_2$ (right) on the $24^3$, $m_la=0.005$ ensemble.}
  \label{fig:R1R2}
\end{figure*}

\begin{figure*}[!ht]
  \centering
  \includegraphics[angle=0,width=0.75\linewidth ]{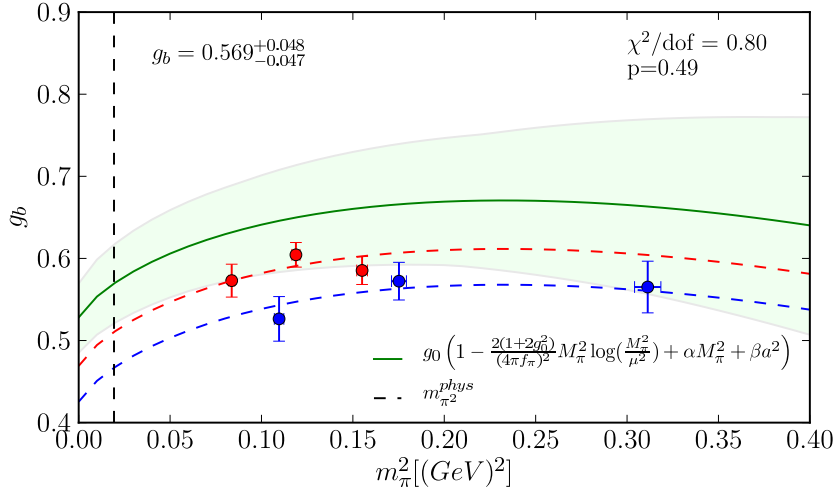}
  \caption{Chiral and continuum extrapolation. The bottom (blue) dashed line is the fit through the $24^3$ ensemble points. The dashed line above (red) is the fit through the $32^3$ ensemble points and the green solid line is the continuum extrapolation with a shaded error band. The intersect with the vertical dashed line corresponds to the physical pion mass. }\label{chiral_extrap}
\end{figure*}

Figure~\ref{fig:R1R2} shows the ratios $R_1$ and $R_2$ on the $24^3$, $m_la=0.005$ ensemble fitted to a constant with statistical errors estimated using single elimination jack-knife. 
We perform a chiral extrapolation using the SU(2) \hmchipt{} formula for the axial coupling matrix element derived in~\cite{Detmold2011}:
\begin{equation}
  \label{eq:BecirevicChiralII}
  g_b = g_0\left(1 - \frac{ 2(1+2g_0^2) }{ (4\pi f_{\pi})^2 } M_{\pi}^2 \log \frac{ M_{\pi}^2 }{ \mu^2 } + \alpha M_{\pi}^2 + \beta a^2\right),
\end{equation} 
which is next-to-leading order in the chiral expansion, but only leading order in the heavy-quark expansion.
We parameterize the light-quark and gluon discretisation effects with an $a^2$ term, as expected for the domain-wall light-quark and Iwasaki gauge actions. The lattice-spacing dependence from the RHQ action is more complicated. However, we estimate the heavy-quark discretisation effects using power counting arguments~\cite{Oktay2008} and find them to be negligible, such that extrapolating in $a^2$ captures the leading scaling behaviour. 
Figure~\ref{chiral_extrap} shows the chiral-continuum extrapolation to the physical light-quark mass and continuum using Eq.~\eqref{eq:BecirevicChiralII}. 
We consider systematic errors arising from uncertainty in the lattice scale, the difference of our sea strange quark mass from its physical value, and uncertainties propagated through from the tuning of the RHQ paramaters. However, we find our dominant source of uncertainties to be the combined chiral and continuum extrapolation. To estimate this error we tried a number of variations to our fitting procedure which included dropping the heaviest masses from each ensemble, considering a linear fit, and a fit function with no lattice scale dependence. We also varied the value of $f_{\pi}$ that appears in Eq.~\eqref{eq:BecirevicChiralII} to simulate changing the relative sizes of the NLO and NNLO terms that appear in the chiral expansion. Our overall estimate of the uncertainty arising from the chiral and continuum extrapolations is 10\%, and adding this in quadrature to the systematic errors from all other sources we arrive at a total error of 10.4\%. Our final value of the $g_b$ coupling including statistical and systematic errors is:
\begin{equation}
  g_b = 0.567(52)_{stat}(59)_{sys}.
\end{equation}
A publication with full details of our analysis and error estimates is in progress.
\section{Conclusions}

The determination of physical quantities from lattice-QCD simulations with unphysically heavy up- and down-quark masses requires a chiral extrapolation to the physical point.  For heavy-light mesons, theoretical guidance is provided by heavy-meson chiral perturbation theory.  At leading order the HM$\chi$PT Lagrangian has one low-energy constant $g$, which we have calculated for the theory with heavy $b$-quarks. Our calculation is the first directly at the physical $b$-quark mass, and has a complete systematic error budget. Comparing our result with other determinations \cite{Ohki2008,Detmold2012, Becirevic2012} shows it lies in the region that would be expected from interpolating between the charm- and infinite-mass.  Our result will be used by the RBC/UKQCD collaboration in the chiral extrapolations of numerical lattice data for the $B$-meson leptonic decay constants \cite{Witzel2012} and $B\to\pi\ell\nu$ semileptonic form factor \cite{Witzel}, and can also be used by other lattice collaborations working on $B$-physics.  This will help to reduce the important, and in many cases dominant, systematic uncertainty from the chiral extrapolation.

\bibliography{/Users/bts105/Documents/Mendeley/library}
\end{document}